\documentclass[prd,aps,twocolumn,a4paper,floatfix,showpacs,nofootinbib]{revtex4-1}

\usepackage{graphicx,psfrag} 
\usepackage{mathrsfs}
\usepackage{amsmath,amsfonts,amssymb} 
\usepackage{multirow}
\usepackage{bm} 
\usepackage{comment} 
\usepackage{hyperref}
\usepackage{placeins} 
\usepackage[normalem]{ulem} 
\usepackage{booktabs} 
\usepackage{xspace}
\usepackage{enumerate}

\def\GMc2{{\rm G M_{\odot} c^{-2}}}

\newcommand{\be}{\begin{equation}} 
\newcommand{\ee}{\end{equation}}
\newcommand{\bea}{\begin{eqnarray}} 
\newcommand{\eea}{\end{eqnarray}}
\newcommand{\bel}{\begin{align}} 
\newcommand{\eel}{\end{align}}
\newcommand{\bse}{\begin{subequations}}
\newcommand{\ese}{\end{subequations}} 
\newcommand{\vv}[1]{\bm{#1}}

\newcommand{\Snn}{SLy$^{(\nearrow \nearrow)}$\xspace}

\newcommand{\Soo}{SLy$^{(0 0)}$\xspace}
\newcommand{\Suu}{SLy$^{(\uparrow \uparrow)}$\xspace}
\newcommand{\Sdd}{SLy$^{(\downarrow \downarrow)}$\xspace}

\newcommand{\Sss}{SLy$^{(\swarrow \swarrow )}$\xspace}

\usepackage{color}

\definecolor{cyan}{rgb}{0,0.9,0.9}
\definecolor{orange}{rgb}{0.9,0.5,0}
\definecolor{magenta}{rgb}{1,0,1}
\definecolor{purple}{rgb}{0.8,0.4,0.8}
\definecolor{gray}{rgb}{0.5,0.5,0.5}

\begin{document}

\title{Numerical Relativity Simulations of Precessing Binary Neutron Star Mergers}

 \author{Tim \surname{Dietrich}$^1$}
\author{Sebastiano \surname{Bernuzzi}$^{2,3}$}
\author{Bernd \surname{Br\"ugmann}$^4$}
\author{Maximiliano \surname{Ujevic}$^5$}
\author{Wolfgang \surname{Tichy}$^6$}

\affiliation{${}^1$Max Planck Institute for Gravitational Physics (Albert Einstein Institute), Am M\"uhlenberg 1, Potsdam-Golm, 14476, Germany}
\affiliation{${}^2$ Dipartimento di Scienze Matematiche Fisiche ed Informatiche, Universit\'a di Parma, I-43124 Parma, Italia}
\affiliation{${}^3$ Istituto Nazionale di Fisica Nucleare, Sezione Milano Bicocca, gruppo collegato di Parma, I-43124 Parma, Italia}
\affiliation{${}^4$ Theoretical Physics Institute, University of Jena, 07743 Jena, Germany}  
\affiliation{${}^5$ Centro de Ci\^encias Naturais e Humanas, Universidade Federal do ABC,09210-170, Santo Andr\'e, S\~ao Paulo, Brazil}
\affiliation{${}^6$ Department of Physics, Florida Atlantic University, Boca Raton, FL 33431 USA}

\date{\today}

\begin{abstract}
We present the first set of numerical relativity simulations of binary neutron
mergers that include spin precession effects and are evolved with multiple resolutions. 
Our simulations employ consistent initial data in general relativity with different spin
configurations and dimensionless spin magnitudes $\sim 0.1$.
They start at a gravitational-wave frequency of $\sim392$~Hz and cover
more than $1$ precession period and about 15 orbits up to merger.  
We discuss the spin precession dynamics by analyzing coordinate
trajectories, quasi-local spin measurements, and energetics, by comparing spin aligned,
antialigned, and irrotational configurations. Gravitational waveforms
from different spin configuration are compared by calculating the
mismatch between pairs of waveforms in the late inspiral.
We find that precession effects are not distinguishable
from nonprecessing configurations with aligned spins for 
approximately face-on binaries,
while the latter are distinguishable from
a nonspinning configurations. Spin precession effects 
are instead clearly visible for approximately edge-on binaries. 

For the
parameters considered here, precession does not 
significantly affect the characteristic postmerger
gravitational-wave frequencies nor the mass ejection. 
Our results pave the way for the modeling of spin precession effects in
the gravitational waveform from binary neutron star events. 
\end{abstract}


\maketitle

\section{Introduction}
\label{sec:intro}
 
The recent observation of gravitational waves (GW) and electromagnetic (EM)
signals from the merger of two neutron stars (NSs) marks a breakthrough 
in the field of multi-messenger 
astronomy~\cite{TheLIGOScientific:2017qsa,GBM:2017lvd}. 
In order to interpret the detected GW and EM signals 
accurate models for the coalescence of compact binaries are required. 
Because of the complexity of Einstein's field equations 
coupled to the equations governing general relativistic hydrodynamics 
those models are based on or compared to 
numerical relativity (NR) simulations. 

To be prepared for future GW detections of unknown 
binary neutron star (BNS) systems, 
one needs to cover the entire parameter space, 
including a variety of Equations of State (EOSs), 
different mass ratios, and NS spins.  
While observations of pulsars in BNS systems suggest 
that most NSs have relatively small spins
\cite{Kiziltan:2013oja,Lattimer:2012nd}, this
conclusion might be biased by the small
number of observed BNSs. 
Indeed, pulsar observations indicate that NSs in binary systems
can have a significant amount of spin. Some stars can even  
approach the rotational frequency of isolated millisecond pulsars.
For example, the NS in the binary system PSR J1807$-$2500B has a
rotation frequency of $239$Hz \cite{Lorimer:2008se,Lattimer:2012nd}. 
In known double NS systems the fastest spinning pulsar, PSR J0737$-$3039A, 
has rotational frequency of $44$Hz \cite{Burgay:2003jj}. 
Assuming magnetic dipole and gravitational wave radiation 
PSR J0737$-$3039A will only spin-down marginally before its merger 
showing that NSs can have a significant amount of spin when they 
merge~\cite{Tichy:2011gw,Dietrich:2015pxa}.

In addition to the spin magnitude, also the orientation of spins 
in BNS systems is highly uncertain. Considering BNSs
formed in situ, misaligned spins with 
respect to the orbital angular momentum 
can be caused by kicks created during the supernova explosions forming the 
two NSs. While after its formation the primary NS might accrete material from the 
secondary star, which is at this time in an earlier stage 
of its evolution and not a NS, this is impossible for the secondary star~\cite{Chruslinska:2017odi}. 
Consequently a realignment of the spin caused by accretion is only likely for the more massive NS.
Thus, the supernova explosion of the secondary star might 
introduce a spin kick creating 
a large misalignement angle. 
For example PSRJ0737-3039B has a spin misaligned with the orbital angular 
momentum by $\approx 130^\circ$~\cite{Farr:2011gs}. 
For BNS systems formed by dynamical capture, e.g.~in globular clusters, 
there is no reason to assume that spins are aligned 
to the orbital angular momentum. 
Overall, further work is needed for a better understanding of the 
formation scenario of BNS systems and 
to quantify the imprint of spin on the binary evolution process.\\

To extract the binary properties, e.g.~spin, from a measured GW signal, 
the detected signal is cross-correlated with template waveforms. 
The requirements for the template bank is twofold.
First, the creation of the waveforms needs to be reasonable 
fast to allow the computation of a large number of waveforms in a small 
amount of time. Second, the templates need to capture accurately the binary evolution. 
For the long inspiral of BNSs detectable by advanced GW interferometers these requirements 
are challenging. 

Considering the recent BNS detection~\cite{TheLIGOScientific:2017qsa} a Post-Newtonian (PN)
approximant~\cite{Sathyaprakash:1991mt,Vines:2011ud} 
and an approximant incorporating results from the effective-one-body model~\cite{Bohe:2016gbl}
with a phenomenological representation of tidal effects~\cite{Dietrich:2017aum} have been used. 
Both models describe binaries in which 
the spins are aligned/anti-aligned with the orbital angular momentum and precession does not occur. 
Thus, an important scientific target is the development of waveform models for 
BNS that include precession.

Also in NR simulations spin and precession effects have not been studied in 
great detail. For a long time spins have been neglected (irrotational binaries)
or have been treated unrealistically by assuming that the stars are tidally locked 
(corotational configurations). 
Only in the last few years, NR groups performed spinning NS simulations
dropping the corotational assumption. 
Most simulations of spinning binaries use approximate initial data by employing constraint violating 
data~\cite{Kastaun:2013mv,Tsatsin:2013jca,Kastaun:2014fna,Kastaun:2016elu} or by 
employing constraint satisfying data which, however, do not
fulfill the equations for hydrodynamical 
equilibrium~\cite{East:2015vix,Paschalidis:2015mla,East:2016zvv}.
Simulations employing constraint solved data fulfilling the 
equations for hydrodynamical equilibrium have been presented 
in~\cite{Bernuzzi:2013rza,Dietrich:2015pxa,Tacik:2015tja,Dietrich:2016lyp,Dietrich:2017aum}.
Most of these simulations focused on spin-aligned cases. 
A precessing NS inspiral has been shown in~\cite{Tacik:2015tja} excluding 
the merger of the stars, and in Ref.~\cite{Dietrich:2015pxa} a single precessing 
simulation at low resolution has been presented. 
The purpose of this article is to present BNS simulations 
for two precessing systems for various resolutions and 
to compare those with spin-aligned configurations. 

The article is structured as follows: in Sec.~\ref{sec:config}
we give an overview of the studied configurations and the employed numerical methods;
in Sec.~\ref{sec:accuracy} accuracy measures for the simulations are presented;
in Sec.~\ref{sec:dynamics} we discuss the dynamics focusing on the precession dynamics 
and the conservative dynamics in terms of binding energy vs.~specific
angular momentum plots; in Sec.~\ref{sec:waves} and Sec.~\ref{sec:ejecta}
we discuss gravitational waves and the dynamical ejecta. 
We conclude in Sec.~\ref{sec:conclusion}.

\section{BNS Configurations}
\label{sec:config}

\begin{table}[t]
  \centering
  \caption{ \label{tab:ID} Initial data for the evolutions considered in this work. 
    The columns refer to: the simulation name,
    the gravitational masses of star A and B, the stars'
    dimensionless angular momenta $\chi^{A,B}$ magnitude and 
    their orientation $\hat{\chi}^{A,B}$. 
    All configurations are evolved with the three different resolutions stated in Tab.~\ref{tab:grid}. 
    This makes a total of 15 simulations.}
  \begin{tabular}{l|cccccccc} 
  \hline
    name & $M^A$& $M^B$ & $\chi^A $& $\chi^B$ & $\hat{\chi}^A$ & $\hat{\chi}^B$ \\ 
    \hline
    \Suu & 1.3547 & 1.1067 & $0.077$ & $0.089$ & $(0,\ 0,\ 1)$    & $(0,\ 0,\ 1)$  \\ 
    \Snn & 1.3553 & 1.1072 & $0.13$  & $0.16$  & $\frac{(1,1,1)}{\sqrt{3}}$& $\frac{(1,1,1)}{\sqrt{3}}$   \\
    \Soo & 1.3544 & 1.1065 & $0.00$  & $0.00$  & ---          & --- \\ 
    \Sss & 1.3553 & 1.1072 & $0.13$  & $0.16$  & - $\frac{(1,1,1)}{\sqrt{3}}$ & - $\frac{(1,1,1)}{\sqrt{3}}$  \\
    \Sdd & 1.3547 & 1.1067 & $0.077$ & $0.089$ & $(0,\ 0,\ -1)$   & $(0,\ 0,\ -1)$  \\     
    \hline
  \end{tabular}
\end{table}

\subsection{Binaries properties}

We study BNSs with two unequal mass NSs
with fixed rest masses of $M_b^A=1.5000M_\odot$
and $M_b^B=1.2000M_\odot$. The configurations differ in their spin magnitudes and spin orientation. 
The NS masses in isolation are $M^A\approx 1.35M_\odot$ and $M^B\approx 1.11M_\odot$, leading to 
a binary mass of $M\approx 2.46M_\odot$. Small differences are
present because of the different spin magnitudes, see details in Tab.~\ref{tab:ID}.

In Ref.~\cite{Dietrich:2015pxa} we have already presented results for one precessing, 
one spin aligned, and one non-spinning configuration employing low resolutions (resolution R1 in Tab.~\ref{tab:grid}). 
Here we include also a simulation for an antialigned spin, and an additional precessing 
configuration. Furthermore, we evolve all systems for three resolutions, 
which allows us to include error estimates.
Details about the initial configurations can be found in Tab.~\ref{tab:ID} and are summarized 
in the following. 

\Snn is the precessing system considered in~\cite{Dietrich:2015pxa}, where 
the two spins of the NSs are pointing along the room diagonal, \Suu
has the same spin magnitude parallel to the orbital angular momentum, which ensures 
for $t=0$ the same spin-orbit contribution as for \Snn. 
However, this leads to a spin magnitude $\sqrt{3}$ times 
smaller than for \Snn. The corresponding non-spinning simulation 
is denoted by \Soo. 
Additionally, we consider \Sss for which the spins are pointing in the 
opposite direction as for 
\Snn and a configuration with spins opposite to \Sdd. 
This ensures initially the same spin-orbit coupling for 
\Sss and \Sdd.

\begin{table}[t]
  \centering    
  \caption{Grid configurations for our simulations. 
    The columns refer to: $L$ total number of levels, 
    $n$ number of points per direction, 
    $L_{\rm mv}$ number of moving box levels using $n_{\rm mv}$ points per direction, 
    $h_0$ coarsest grid spacing, and
    $h_L$ finest grid spacing.} 
  \begin{tabular}{l|ccccccc}        
    \hline
    Name & $L$ &  $n$ & $L_{\rm mv} $ & $n_{\rm mv}$ & $h_0$ & $h_L$ \\
    \hline
    R1    & 7   & 160 & 4 & 64  & 15.68  & 0.245 \\
    R2    & 7   & 240 & 4 & 96  & 11.76   & 0.184 \\
    R3    & 7   & 320 & 4 & 128 & 7.84  & 0.123     \\    
    \hline
  \end{tabular}
  \label{tab:grid}
\end{table}

\subsection{Numerical methods}

The initial data for our numerical simulations are computed with the 
pseudo-spectral code SGRID~\cite{Tichy:2009yr} allowing to 
construct spinning NSs with arbitrary spin and 
different EOSs~\cite{Tichy:2012rp,Dietrich:2015pxa}.
Although SGRID allows the construction of eccentricity reduced intial data, 
we do not perform any kind of eccentricity reduction to save computational costs. 
However, the residual eccentricities in our simulations are small 
and stay below $\lesssim 10^{-2}$. We use the same resolution 
as in our previous works, namely $n_A=n_B=n_{\rm cart}+4 = 28 $ points, 
see~\cite{Tichy:2012rp,Dietrich:2015pxa} for a detailed description.

The dynamical simulations are performed with the finite differencing 
code BAM~\cite{Brugmann:2008zz,Thierfelder:2011yi,Dietrich:2015iva,Bernuzzi:2016pie}.
The numerical method is based on the method-of-lines, Cartesian grids and finite
differencing. The grid is made out of a hierarchy of
cell-centered nested Cartesian boxes consisting of $L$ refinement levels, 
which we label with $l = 0,...,L-1$ ordered by increasing resolution. 
The resolution inside each level increases by a factor of $2$ and can be 
computed according to $h_l= 2^{-l} h_0$. For completeness we give $h_L$ and $h_0$ 
in Tab.~\ref{tab:grid}. The inner levels employ $n_{\rm mv}$ points per direction and move 
following the technique of `moving-boxes', while the outer levels remain fixed and employ 
$n$ grid points per direction. We use a Runge-Kutta type integrator for the time evolution. 
For the time stepping the Berger-Collela scheme is employed
enforcing mass conservation across refinement boundaries~\cite{Berger:1984zza,Dietrich:2015iva}. 
Metric spatial derivatives are approximated by fourth order finite
differences. The general relativistic hydrodynamic equations are solved with standard
high-resolution--shock-capturing schemes based on primitive reconstruction
and the local Lax-Friedrich central scheme for the numerical
fluxes, see~\cite{Thierfelder:2011yi,Bernuzzi:2012ci}.

Simulating spin precession does not allow us to impose any
grid symmetry. This is the major difference to most of our previous
simulations, in which we imposed bitant symmetry, and results in double 
computational cost per configuration.
In order to compare the precessing runs to the non-spinning or
spin-aligned ones, we choose not to impose any symmetry also for the
non-precessing configurations.  
The grid configurations are given explicitly in Tab.~\ref{tab:grid}, 
we use three different resolutions for 
all setups denoted by R1, R2, R3. 

\section{Simulations' accuracy} 
\label{sec:accuracy}

\begin{figure}[t]
\includegraphics[width=0.5\textwidth]{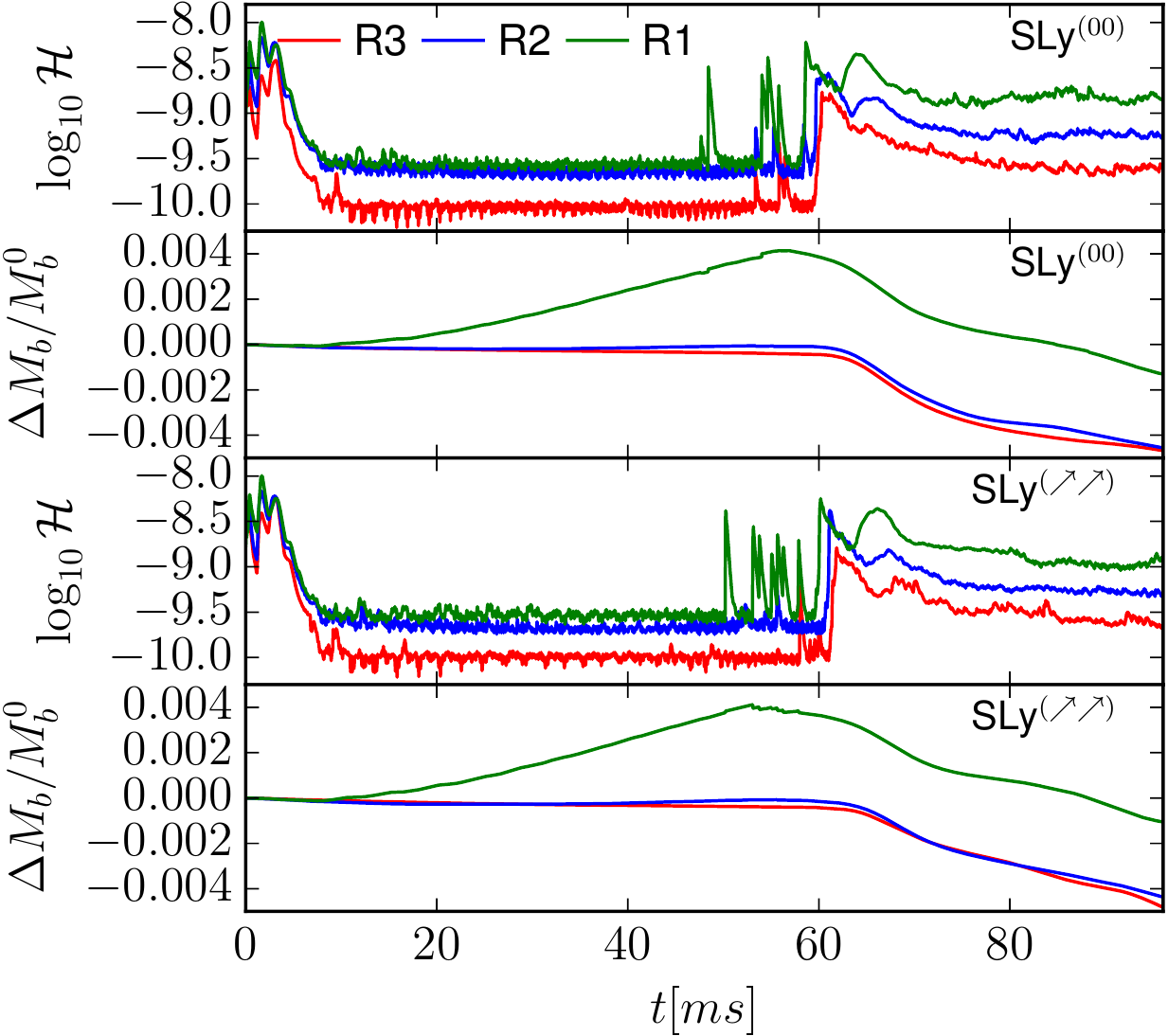}
\caption{ Hamiltonian constraint (first and third panel) 
and restmass conservation (second and fourth panel)
for the non-spinning case \Soo (top) 
and the precessing case \Snn (bottom). 
 }
\label{fig:hamD}
\end{figure}

This section discusses the accuracy of our simulations. Previous
detailed studies have been presented
in~\cite{Bernuzzi:2011aq,Dietrich:2015iva,Bernuzzi:2016pie,Dietrich:2016hky,Dietrich:2017feu}.
This work differs from previous ones mainly because no symmetry
assumption for the computational grid has been made. Note also that our current best production runs employ
high-order finite differencing operators for hydrodynamics
\cite{Bernuzzi:2016pie}, while here we employed a robust but second-order
scheme for the numerical flux. [The higher-order algorithm was not
available when this project was started.]

\subsection{Constraint violation and mass conservation}

In order to assess the validity of our simulations we present the 
$L_2$ volume norm of the Hamiltonian constraint and the 
conservation of rest mass in Fig.~\ref{fig:hamD} for the 
precessing case \Snn (top panels) and for the non-spinning case \Soo
(botton panels). 

Due to the constraint propagation and damping properties of the Z4c
evolution system the constraint stays at or below the value of the initial data.
The main origin of the wiggles during the orbital motion is the mesh
refinement following the motion of the NSs and reflections from the
interface between the shells and Cartesian boxes.  
At merger the constraint grows by about one order of
magnitude due to regridding and to the development of large gradients
in the solution, but it remains below the initial level. Subsequently,
the violation is again propoagated away and damped.
Throughout the simulation we find that the Hamiltonian constraint
violation improves monotonically with increasing resolution.

The violation of the rest-mass conservation happens at mesh refinement
boundaries and due to the artificial atmosphere treatment, 
and possibly mass leaving the computational 
domain. 
Considering the time evolution of the 
mass violation, we find that, independent of the 
spin, the resolution R1 shows an increasing mass during the orbital
motion. That is caused by insufficient resolution 
and the artificial atmosphere treatment. For resolutions R2 and R3 the 
rest mass stays constant within $0.005\%$ during the inspiral. 
After the merger the rest mass is decreasing. The mass loss is caused
by the ejected material which decompresses while it
leaves the central region of the numerical domain. 
Once the density drops by $9$ orders of magnitude, 
the material is counted as atmosphere and not further evolved.
Consequently, conservation of total mass is violated.
Overall the mass violation is below $0.4\%$.

\subsection{Waveform accuracy}

\begin{figure}[t]
\includegraphics[width=0.5\textwidth]{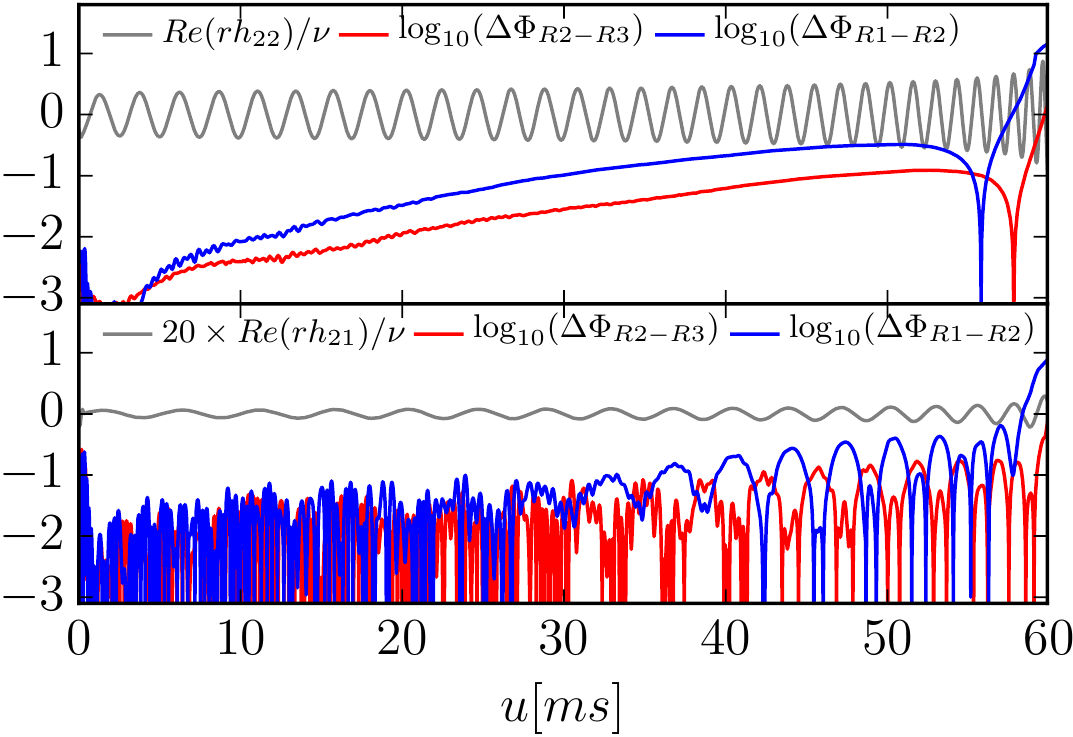}
\includegraphics[width=0.5\textwidth]{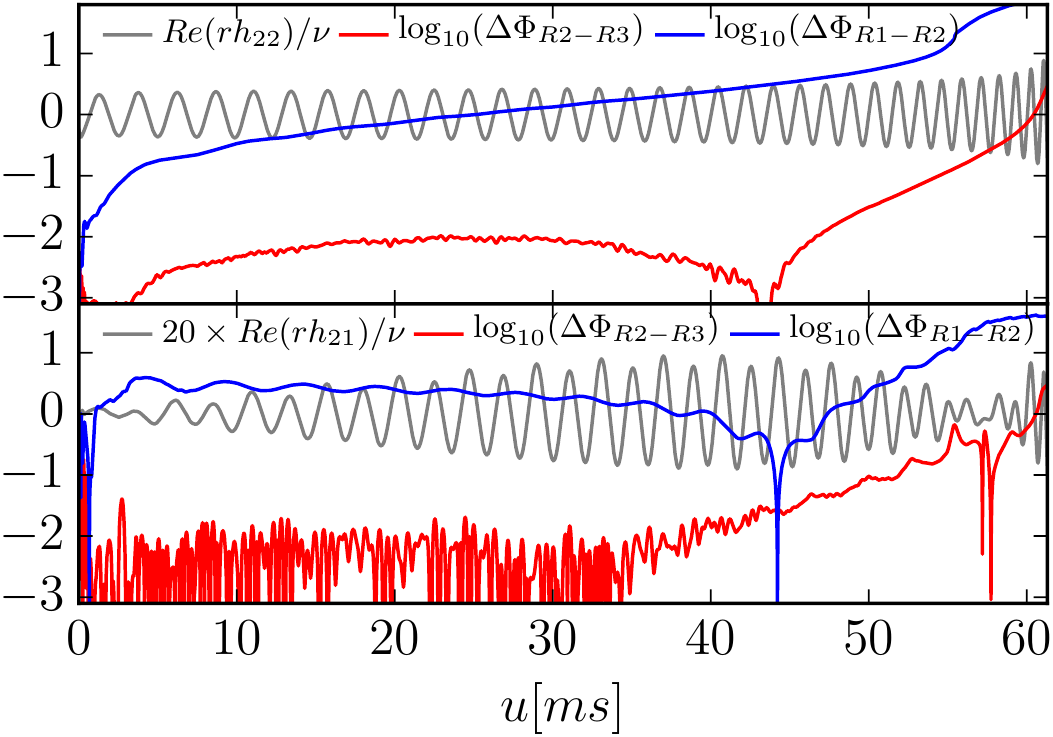}
\caption{
Real part of the (2,2) and (2,1) mode for resolution R3 
as well as the phase difference between different resolutions for the 
\Soo (top) and \Snn (bottom) configurations shown versus retarded time. 
We multiply the amplitude of the (2,1) mode by a factor of $20$ 
for better visibility.}
\label{fig:hconvergence}
\end{figure}

In Fig.~\ref{fig:hconvergence} we present the GW phase difference between different resolutions
for \Soo (top panels) and \Snn (bottom panels) during the inspiral up to the moment 
of merger, which we define as the time of maximum amplitude in the (2,2)-mode.
Through most of the inspiral we see a monotonic decrease of the phase difference for increasing 
resolution, however, a few orbits before merger the phase difference has a zero
and grows again. 
Such zero crossings are caused by competing effects influencing the overall phase evolution, 
in particular, the high numerical viscosity for small resolutions and the 
violation of mass conservation. These zero-crossings have been observed
also in other numerical simulations~\cite{Bernuzzi:2016pie,Kiuchi:2017pte}.   
In~\cite{Bernuzzi:2016pie} and \cite{Dietrich:2017aum} we found that 
employing high-order schemes leads to a constant convergence order 
for the GW phase for multiple EOSs and grid setups. 
A similar convergence order can be achieved also for precessing
binaries if high-order schemes are employed
\cite{Radice:2013hxh,Bernuzzi:2016pie}. While not yet in convergent
regime, our results are consistent and improve with the grid
resolution. 

\section{Dynamics} 
\label{sec:dynamics}

\subsection{Precession Dynamics}

\begin{figure}[t]
\includegraphics[width=0.5\textwidth]{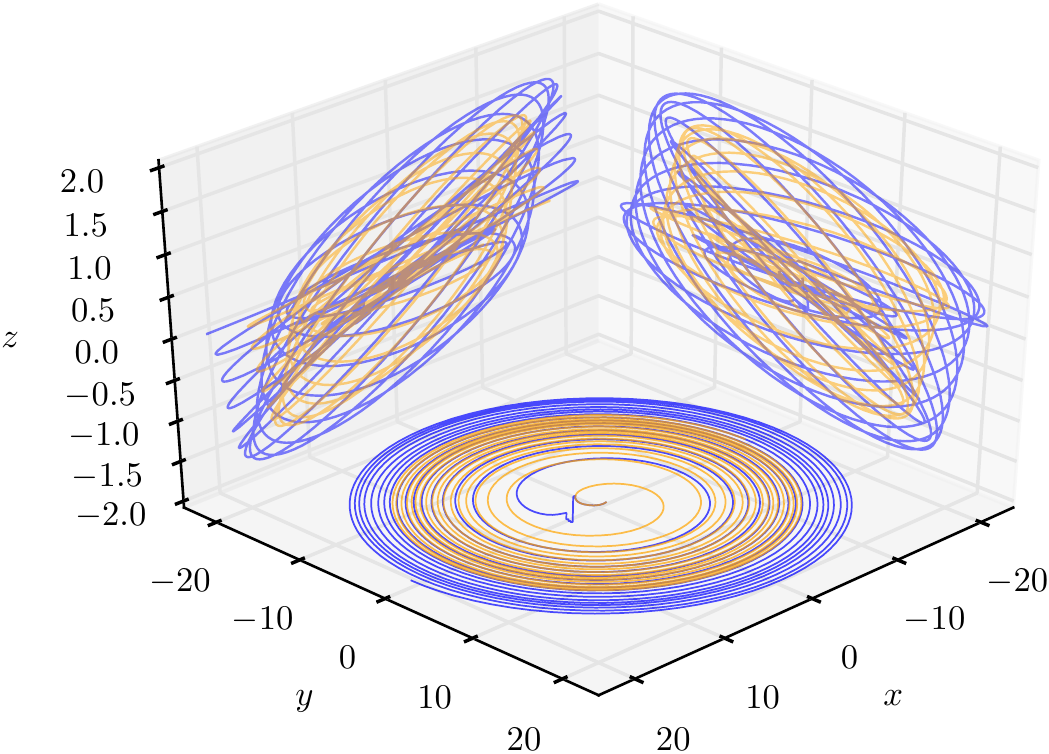}
\includegraphics[width=0.5\textwidth]{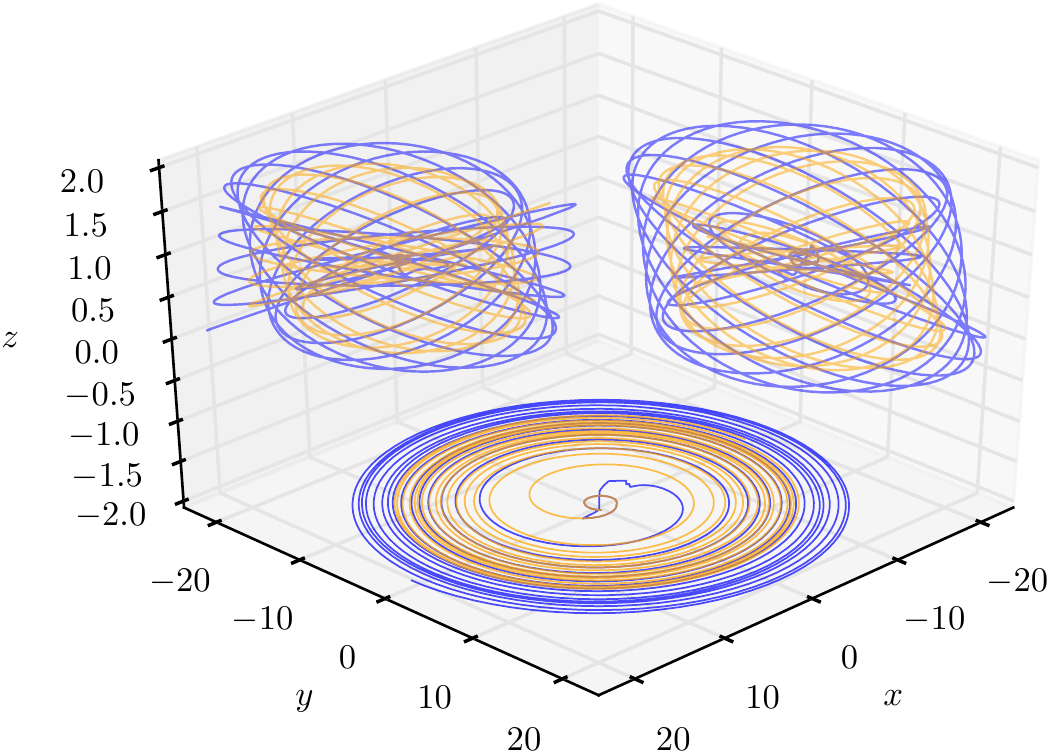}
\caption{
Projection of the NS trajectories of \Snn (top panel) and 
\Sss (bottom panel) onto the $x$-$y$, $x$-$z$, $y$-$z$-plane.
Initially the orbital plane and the $x$-$y$ plane coincide. 
The trajectory of the more massive star $A$ is shown orange and the 
trajectory of its companion (star $B$) is drawn blue.}
\label{fig:tracks}
\end{figure}

All systems start at an initial GW frequency of $M\omega_{22}=0.030$, 
which corresponds to a frequency of $392$Hz. This leads to about 
$\sim 15$ orbits before the NSs merge for the non-spinning case, 
where the exact number 
depends on the spin magnitude and orientation. 
In Fig.~\ref{fig:tracks} we present the trajectories of the NSs for 
\Snn (top panel) and \Sss (bottom panel). 
Initially the two stars are located at ${\bf r}_A=(-17.2,0.0,0.0)$ and ${\bf r}_B=(21.0,0.0,0.0)$. 
They start to leave the $x$-$y$-plane during the 
simulation due to the misaligned spin~\cite{Kidder:1995zr,Apostolatos:1994mx}. 
The maximum angle between the $x$-$y$-plane and the orbital plane can be roughly estimated 
as $\alpha \approx z(t^*)/\sqrt{x^2(t^*)+y^2(t^*)}$, where $t^*$ is chosen in a way such 
that $z$ is maximal. We find $\alpha \approx 0.11 \approx 6^\circ$. 

\begin{figure}[t]
\includegraphics[width=0.5\textwidth]{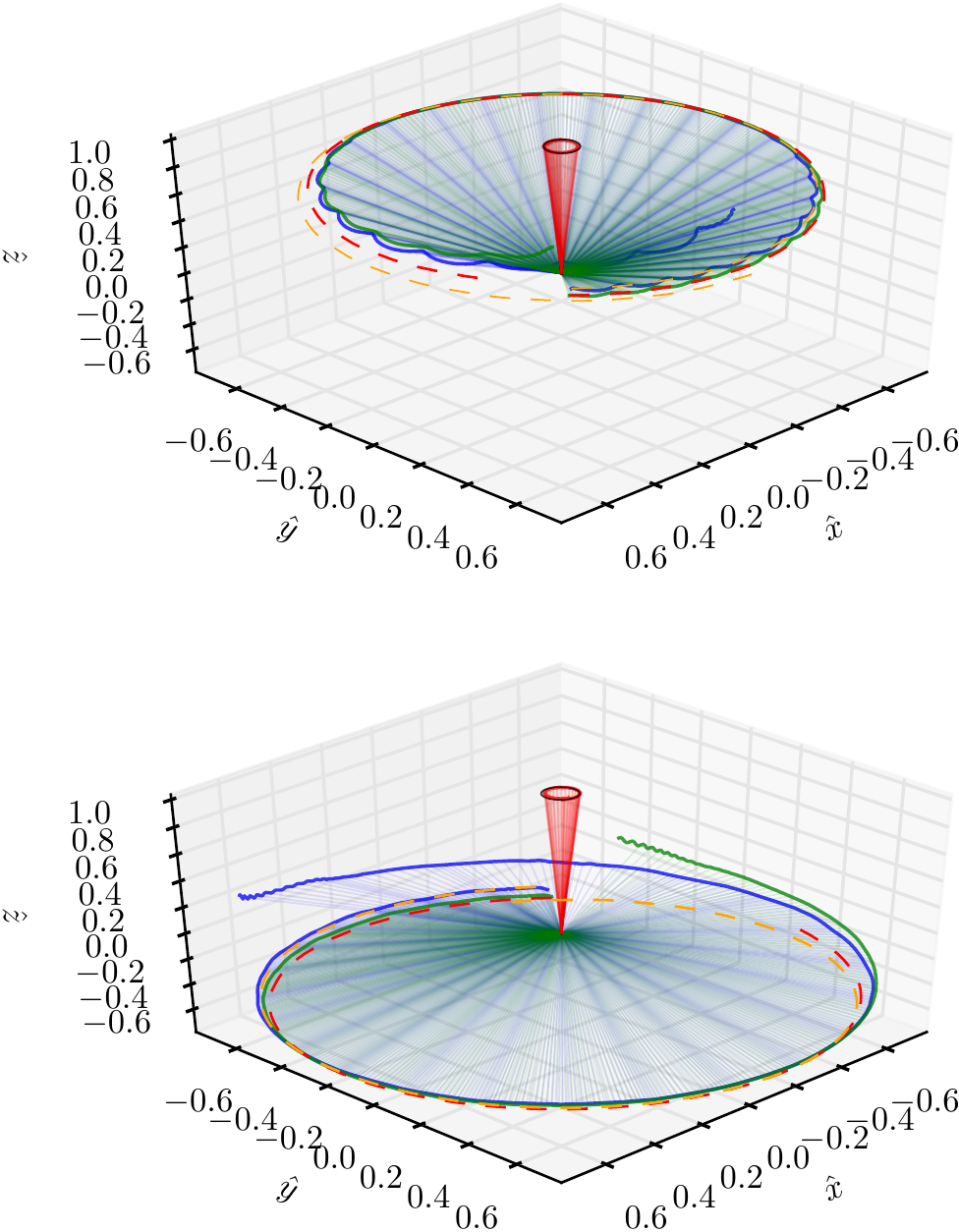}
\caption{Evolution of the orientation of the angular momentum (solid red curve), the 
spin of the primary NS (solid green curve) and the secondary star (solid blue curve) for \Snn (top panel) 
and \Sss(bottom panel).
The individual spin of the stars is estimated according to Eq.~\eqref{eq:quasi_local_spin}
on coordinate spheres with $r_s=14$.
We also include PN estimates obtained with the code of~\cite{Boyle:2013nka}
using the NR data extracted about $1{\rm ms}$ after the start of 
the simulation as initial conditions, where the precession cone of the 
orbital angular momentum is shown as a think black line, of the spin of the primary star as a dashed red line 
and of the secondary star as a orange dashed line. }
\label{fig:cones}
\end{figure}

We further access the precession dynamics by computing the precession cones, cf.~Fig.~\ref{fig:cones}. 
The figure shows the precession cones for the orbital angular momentum (red), the spin of star A (green), 
and the spin of star B (blue) for the configurations \Snn (top) and \Sss (bottom). 
The precession cones for the individual stars are based on 
the quasi-local spin measurement introduced in~\cite{Dietrich:2016lyp}
\begin{equation}
\label{eq:quasi_local_spin}
 S^i \approx \frac{1}{8\pi} 
 \int_{r_s} \text{d}^2 x \sqrt{\gamma} \left( \gamma^{k j} K_{l k} - \delta_{l}^{j} K \right) 
 n_{j} \varphi^{li}, 
\end{equation}
evaluated on coordinate spheres with radius $r_s$ around the NSs. 
$\varphi^{li} = \epsilon^{l i k } x_k $ defines the approximate rotational Killing vectors 
in Cartesian coordinates ($\varphi^{l1},\varphi^{l2},\varphi^{l3}$), 
$K_{ij}$ denotes the extrinsic
curvature, $\gamma^{ij}$ is the inverse 3-metric,  
and $n_i = (x_i-x_i^{\rm NS})/r$ is the normal vector with respect to the
center of the NS, see also \cite{Tacik:2015tja,Kastaun:2016elu}.
The precession cone considering the orbital angular momentum is 
computed from the stars' trajectories, where we use 
$\vv{\hat{L}} = \vv{\hat{r}} \times \text{d}\vv{\hat{r}}/\text{d}t$, 
with $\vv{r}$ being the vector connecting 
star A and B.

From Fig.~\ref{fig:cones} one can see that both configurations \Sss and \Snn 
contain more than one full precession cycle. In addition to precession, 
the system also undergoes nutation, i.e.~small oscillations 
occurring at about twice the orbital frequency and thus 
on a timescale much shorter than the precession timescale.
These nutation cycles are visible for the individual spins for \Snn, 
but are also present for \Sss. Caused by the particular definition of the 
spin axis nutation effects in NR simulations differ from those in PN 
theory~\cite{Owen:2017yaj}.

We compare the numerical relativity results with PN predictions,
obtained from the TaylorT4 approximant as implemented in 
\texttt{GWframes}~\cite{Boyle:2013nka}. To allow a direct comparison we use as 
initial conditions the spin and frequency estimated about $1{\rm ms}$ after 
the beginning of the simulation. 
We find agreement between the NR
results and the PN predictions regarding the overall spin precession.
However, during the late inspiral the PN and NR data start to disagree. 
We surmise that 
the differences are caused by (i) the fact that the PN prediction looses 
its accuracy close to the merger; (ii) that possible tidal effects 
start to effect the evolution; (iii) that the quasi-local spin measure 
has a decreasing accuracy once the two stars approach each other. 

\begin{figure}[t]
\includegraphics[width=0.5\textwidth]{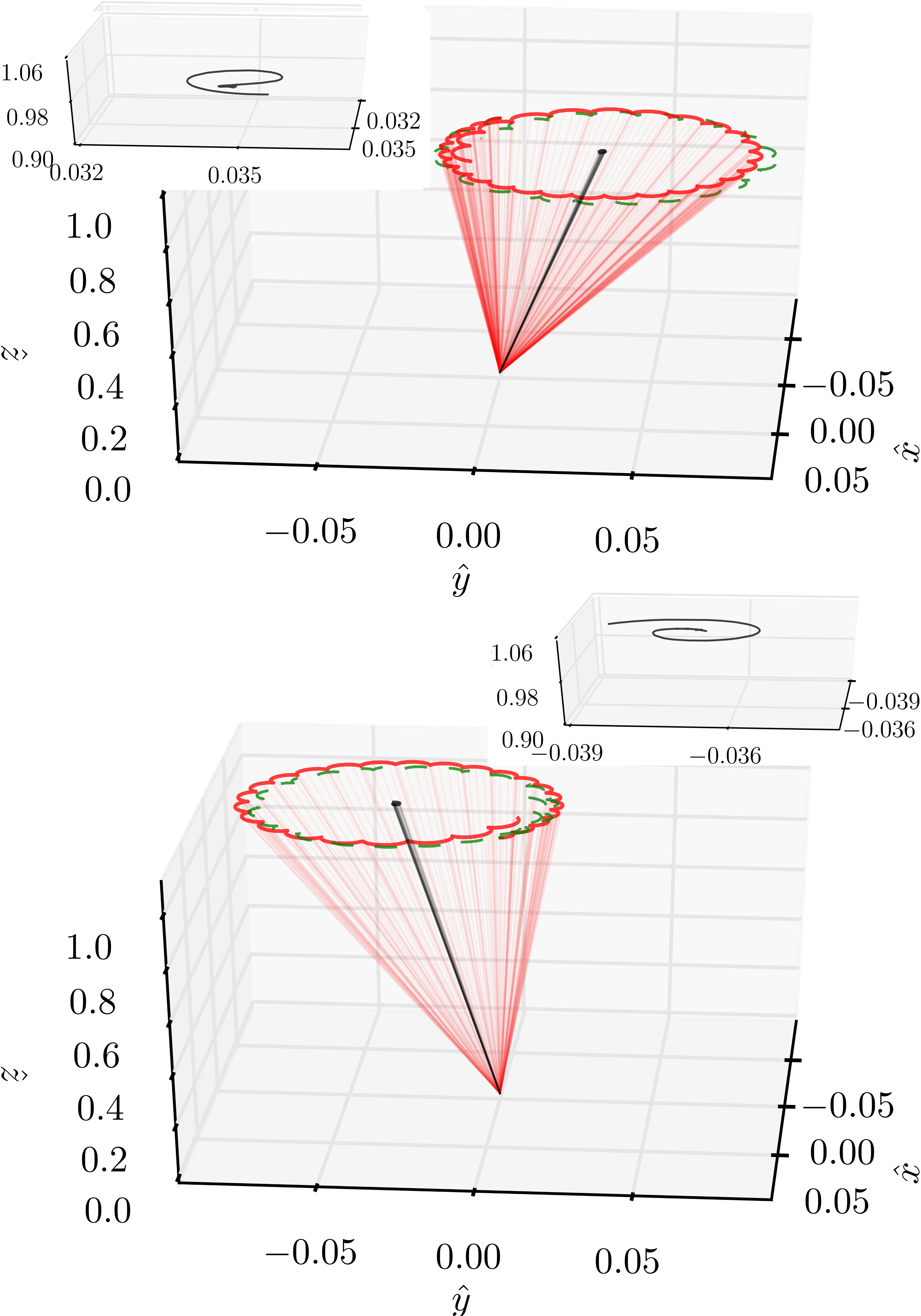}
\caption{Precession cone for \Snn (top) and \Sss configurations (bottom) for the orbital angular momentum 
(red). Additionally we present the precession cone of the total angular momentum black, 
see insets above the main panels.
As green dashed line we show the precession cones for $(-\hat{L}_x,-\hat{L}_y,\hat{L}_z)$ for 
\Sss in the top and \Snn in the bottom panel.}
\label{fig:nutation}
\end{figure}

Let us further focus on the evolution of the orbital angular momentum (red) and 
total angular momentum (black) for the precessing systems, see Fig.~\ref{fig:nutation}. 
While the orbital angular momentum is estimated as before from the tracks of the individual stars, 
we use for the total angular momentum the ADM-angular momentum of the simulated spacetime.
From the figure we can conclude that:
(i) The total angular momentum (black), estimated from the ADM angular momentum $J_{\rm ADM}$ does 
not coincide with the $z$-direction of the numerical grid due to the intrinsic spin of the NSs. 
(ii) The total angular momentum (black) is not constant due to the emission of GWs 
(see inset in Fig.~\ref{fig:nutation}). 
(iii) In addition to the precession of the orbital angular momentum, 
nutation is visible in the orbital angular momentum.
Precession and nutation are expected to occur in such systems because the
total angular momentum vector is not aligned with any of the principal axes
of the moment of inertia tensor of the system.

In Fig.~\ref{fig:nutation} we also present a consistency check based
on the symmetry of \Sss and \Snn. 
The green dashed lines show for each
of the two configurations the precession cone obtained by flipping
$(\hat{L}_x,\hat{L}_y,\hat{L}_z) \rightarrow
(-\hat{L}_x,-\hat{L}_y,\hat{L}_z)$ the angular momentum of the other
configuration (\Sss in the top and \Snn in the bottom panel). Under 
this transformation the precession cones for \Sss and \Snn agree very well, 
if the sign of the $x$- and $y$-components is flipped. 
Note that this consistency check only works if the spins of the neutron stars 
are considerably smaller than the orbital 
angular momentum, since it requires $L_z \gg S_z$, 
since otherwise also an adjustment of $\hat{L}_z$ needs to be taken into account. 

\subsection{Binding energy curves}

\begin{figure}[t]
\includegraphics[width=0.5\textwidth]{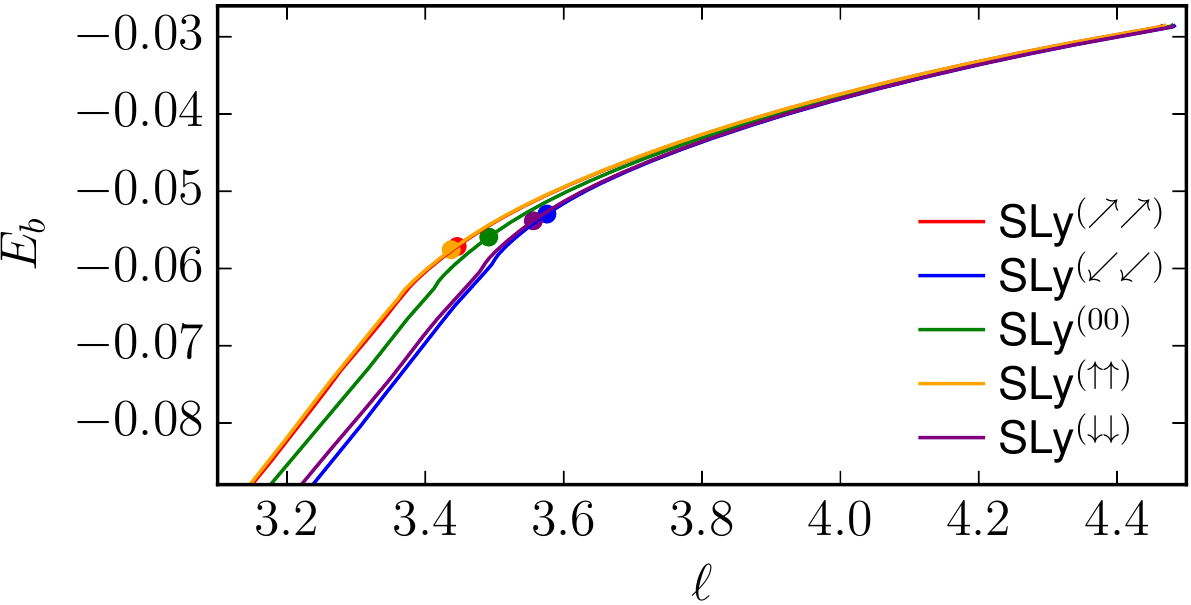}
\caption{Binding energy $E_b$ as a function of the specific orbital angular momentum $\ell$ 
for all configurations considered in the article. 
Circles represent the moments of merger. }
\label{fig:Ej}
\end{figure}

\begin{figure}[t]
\includegraphics[width=0.5\textwidth]{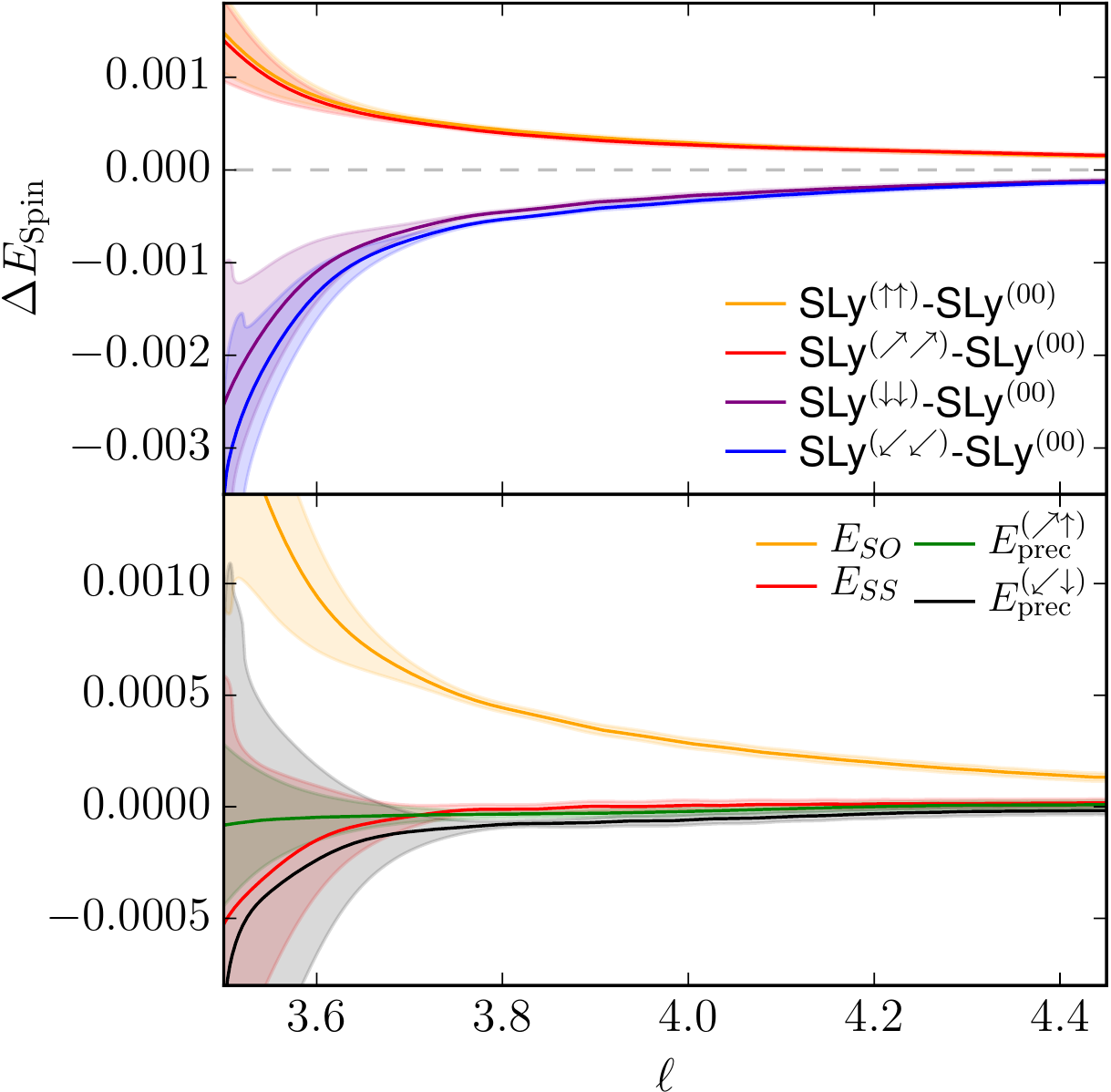}
\caption{Top panel: Estimate of the spin effects on the conservative dynamics by 
taking the difference between cases including spin and the irrotational configuration. 
The shaded region marks the difference to results obtained with a lower resolution 
and takes into account the uncertainty of the initial data. 
Bottom panel: Contributions to the binding energy estimated following the discussion 
given in the text.}
\label{fig:Ej_contr}
\end{figure}

In addition to the qualitative discussion presented above, we also 
investigate the conservative dynamics constructing the binding energy 
\begin{equation}
E_{b}=\frac{M_{\rm ADM}(t_0)-E_{\rm rad}-M}{\nu \ M}, 
\end{equation}
and the reduced orbital angular momentum 
\begin{equation}
\ell =\frac{|{\bf{J}_{\rm ADM}}(t_0)- {\bf{S}_1}(t_0) -{\bf{S}_2}(t_0) -\bf{J}_{\rm rad}|}{\nu\ M^2}. 
\end{equation}
Where $\nu=M^A M^B/M$ is the symmetric mass ratio, 
$E_{\rm rad},\bf{J}_{\rm rad}$ are the emitted energy and angular momentum due to GWs, 
and $M_{\rm ADM}, J_{\rm ADM}$ denote the 
ADM-mass and angular momentum at the beginning of the simulation ($t = t_0$), 
see e.g.~\cite{Damour:2011fu,Bernuzzi:2013rza}. 
${\bf{S}_1}(t_0)$ and ${\bf{S}_2}(t_0)$ are estimated from the initial data, 
Tab.~\ref{tab:ID}, and ${\bf{S}}_{A,B} = M^{A,B} \chi^{A,B} \hat{\chi}^{A,B}$.

Figure~\ref{fig:Ej} shows the $E$-$\ell$ curve 
for all configurations employing the highest resolution R3. 
During the inspiral (large $\ell$ and $E$) the curves are almost 
indistinguishable. When the stars approach each other due to the emission of 
energy and angular momentum, we see clear differences between 
configurations with aligned spin, without spin, and with anti-aligned spin.
This becomes even more prominent in the top panel of Fig.~\ref{fig:Ej_contr}
in which we show the difference between all spinning cases with respect to the 
non-spinning configuration \Soo. 
The shaded region represents our error estimate, which we obtain from 
assigning to every configuration an error due to the finite resolution 
estimated by taking the difference between resolution R2 and R3 and 
by adding an additional uncertainty of $ 10^{-5}$ to account for the 
uncertainty of the initial data solver~\cite{Dietrich:2015pxa}. 
The final error of the linear combination 
is obtained from error propagation assuming errors of different 
configurations are uncorrelated. Note that in case one estimates the error directly from the 
linear combinations obtained for different configurations, 
the error reduces by about a factor of $3$, thus 
we suggest that the error estimates shown in Fig.~\ref{fig:Ej_contr}
are conservative estimates. 

Assuming that the total binding energy consists of a non-spinning
contribution including tidal effects $E_0$, 
a spin-orbit contribution $E_{SO}$, and a spin-spin contribution $E_{\rm SS}$, we 
can write 
\begin{equation}
E_b = E_0 + E_{SO} + E_{SS} + \mathcal{O}(S^3), \label{eq:ansatzEb}
\end{equation}
see also \cite{Bernuzzi:2013rza,Dietrich:2016lyp}.

The bottom panel of Fig.~\ref{fig:Ej_contr} shows the different contributions 
to the binding energy. The spin-orbit term is estimated according to 
\begin{equation}
 E_{SO} = \frac{E_b[{\rm SLy}^{(\uparrow \uparrow)}] - E_b[{\rm SLy}^{(\downarrow \downarrow)}]}{2}
\end{equation}
and in general is at leading order proportional to 
$(\propto\vv{L}\cdot \vv{S}_i/r^3)$, see~\cite{Kidder:1992fr}. 
The spin-spin term consists of a self-spin term and an interaction term between the two spins. 
The interaction term is given by 
$E_{SS} \propto [3(\vv{n}\cdot\vv{S_1})(\vv{n}\cdot\vv{S}_2)-(\vv{S}_1\cdot\vv{S}_2)]/r^3$, 
see e.g.~\cite{Kidder:1992fr}. 
For the precessing systems the spin-spin contribution at the beginning of the simulation 
is equal to zero due to the particular choice of the spin orientation. 
We approximate the full spin-spin term (including spin-spin interaction and self-spin) by 
\begin{equation}
 E_{SS} = \frac{E_b[{\rm SLy}^{(\uparrow \uparrow)}] + E_b[{\rm SLy}^{(\downarrow \downarrow)}]}{2}- 
 E_b[{\rm SLy}^{(0 0)}].
\label{eq:ESS}
 \end{equation}
From Fig.~\ref{fig:Ej_contr} one sees that the spin-spin contribution acts attractive for aligned spin, 
but is relatively small during most of the inspiral and hardly resolved in our simulations. 

To understand the influence of precession, we compute the differences: 
\begin{eqnarray}
 E_{\rm prec}^{(\nearrow \uparrow)} & = & E_b[{\rm SLy}^{(\nearrow \nearrow)}]- 
 E_b[{\rm SLy}^{(\uparrow \uparrow)}],  \\
 E_{\rm prec}^{(\swarrow \downarrow)} & = & E_b[{\rm SLy}^{(\swarrow \swarrow)}]- 
 E_b[{\rm SLy}^{(\uparrow \uparrow)}].   
\end{eqnarray}

We find overall that $E_{\rm prec}^{(\nearrow \uparrow)}$ and 
$E_{\rm prec}^{(\swarrow \downarrow)}$ are consistent with zero within our 
conservative error estimate. Consequently, the conservative dynamics 
shows only minor differences between precessing systems 
and systems with the same effective spin but purely aligned/anti-aligned spin. 
Furthermore, also the spin-spin contribution is within our conservative 
uncertainty estimate consistent with zero. 

\section{Gravitational Waves}
\label{sec:waves}

\begin{figure*}[t]
\includegraphics[width=1.\textwidth]{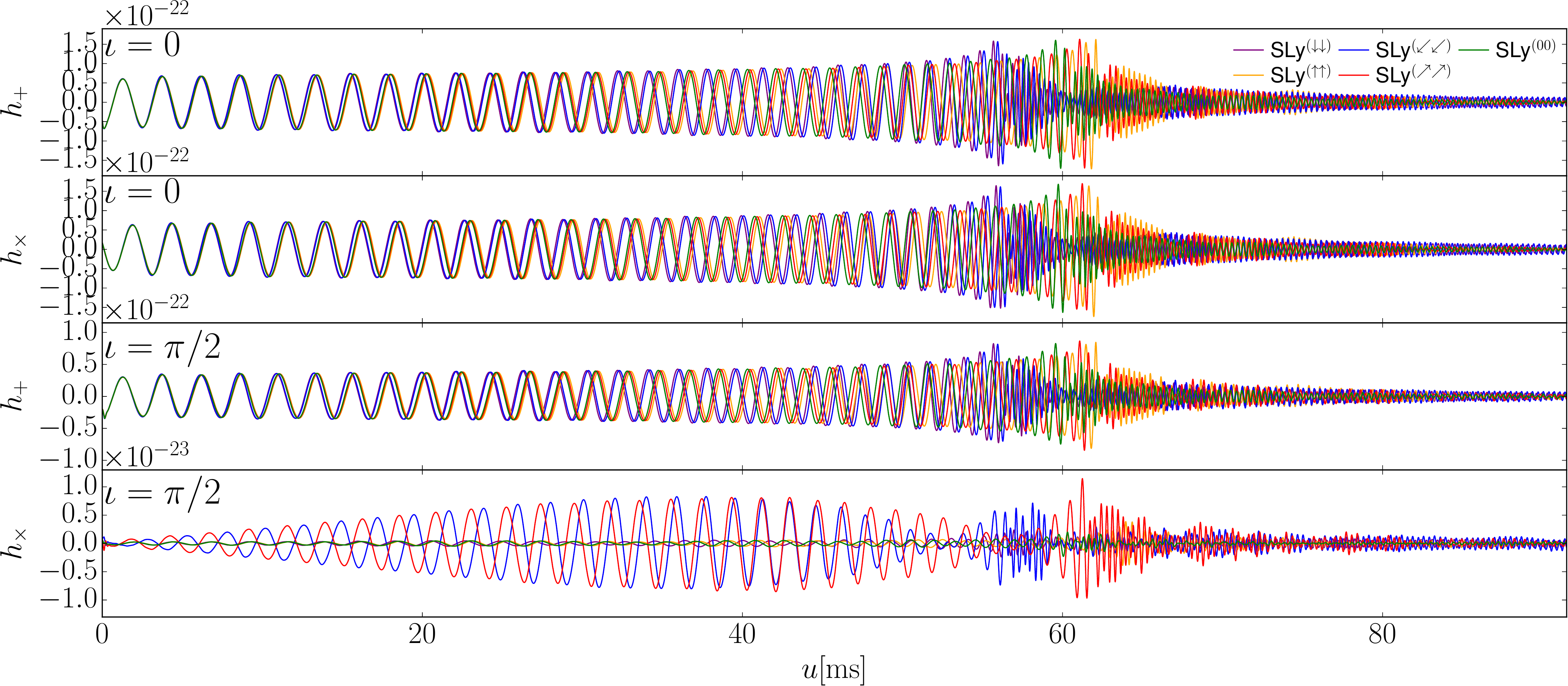}
\caption{Gravitational wave strains $h_+$ (first and third panel) 
and $h_\times$ (second and fourth panel) for the inclinations 
$\iota = 0$ (face on, top panels) and $\iota=\pi/2$ 
(edge on, bottom panels).   
We assume a distance to the binary systems of $100{\rm Mpc}$.}
\label{fig:h}
\end{figure*} 

\subsection{Inspiral Waveform} 
 
\subsubsection{Qualitative Discussion} 
 
We extract GWs from the curvature invariant $\Psi_4$ 
projected on the spin-weighted spherical harmonics for 
spin $-2$, ${}^{-2}Y_{lm}$, see e.g.~\cite{Brugmann:2008zz}. 
Metric multipoles $r h_{\ell m}$ are reconstructed from the 
curvature multipoles using the frequency domain
integration of~\cite{Reisswig:2010di}. 
To construct the GW strain $h$ we sum 
all modes up to $l\leq 4$. 
All waveforms are shown versus the retarded time computed as 
\begin{equation}
 u=t-r_*=t-r_{\rm extr}-2M\ln\left(r_{\rm extr}/2M-1\right).
\end{equation}

In Fig.~\ref{fig:h} we present $h_+$ and $h_\times$, 
\begin{equation}
h_+ - {\rm i} h_\times = \sum_{\ell=2}^{4} \sum_{m = -\ell}^{\ell} h_{\ell
  m}{}^{-2}Y_{\ell m} (\theta = \iota,\phi =0) \ ,
\end{equation}
for two inclinations: face on $\iota=0$ (two top panels) and 
edge on $\iota=\pi/2$ (two bottom panels). 
According to the plot the following observations can be made: 
\begin{enumerate}[(i)]
\itemsep-1.mm 
\item spin-aligned systems merge later, spin anti-aligned systems 
earlier. This effect, also referred to as orbital hang-up 
effect~\cite{Campanelli:2006uy}, is caused by the interaction between 
spins and the orbital angular momentum \cite{Damour:2001tu}. 
\item for $\iota=0$ no imprint of precession is visible 
and cases with the same effective spin evolve similarly. 
\item the amplitude of the GWs for $\iota=\pi/2$ 
is for h$_+$ about a factor of 2 and for $h_\times$
more than an order of magnitude smaller than for $\iota =0$.
\item precession effects with more than one precession cycle are visible 
in $h_\times$ for $\iota=\pi/2$. 
The amplitude of $h_\times$ ($\iota=\pi/2$) 
for the non-precessing systems is significantly smaller than for the 
precessing systems, but non-zero due to the unequal 
masses of the two stars. 
\end{enumerate}

\subsubsection{Mismatch}

\begin{table}[t]
  
\caption{Mismatch between the different configurations for 
$h_+$ (upper triangle of the table) and $h_\times$ (lower triangle of the table).  
We assume two different inclinations $\iota=0$ (top), $\iota = \pi/2$ (bottom). 
We restrict our analysis to a frequency window of $f \in [450,2048]$ and
use the \texttt{ZERO\_DET\_high\_P} noise curve of~\cite{Sn:advLIGO} for the computation 
of the mismatch.}
  \label{tab:precession}   
  
$\iota = 0$

\begin{tabular}{c|ccccc}      
  $h_+/h_\times$  & $(\downarrow \downarrow )$ & $(\swarrow \swarrow)$ & $(00)$  & $(\nearrow \nearrow)$ &  $(\uparrow \uparrow)$ \\ 
\hline 
$(\downarrow \downarrow)$ & -      & 0.0041 & 0.0633 & 0.1179 & 0.1182 \\
$(\swarrow \swarrow)$     & 0.0032 &    -   & 0.0771 & 0.1209 & 0.1217 \\
$(00)$                    & 0.0593 & 0.0361 &    -   & 0.0381 & 0.0416 \\
$(\nearrow \nearrow)$     & 0.1137 & 0.1179 & 0.0361 &   -    & 0.0025 \\
$(\uparrow \uparrow)$     & 0.1138 & 0.1186 & 0.0390 & 0.0007 &   -    \\
\end{tabular}

\vspace*{0.8cm}

$\iota = \pi/2$  

\begin{tabular}{c|ccccc}      
  $h_+/h_\times$  & $(\downarrow \downarrow )$ & $(\swarrow \swarrow)$ & $(00)$  & $(\nearrow \nearrow)$ &  $(\uparrow \uparrow)$  \\ 
\hline 
$(\downarrow \downarrow)$ & -      & 0.0043 & 0.0609 & 0.1166 & 0.1174 \\
$(\swarrow \swarrow)$     & 0.3762 &    -   & 0.0735 & 0.1207 & 0.1209\\
$(00)$                    & 0.1484 & 0.3549 &    -   & 0.0434 & 0.0460 \\
$(\nearrow \nearrow)$     & 0.3512 & 0.0592 & 0.2765 &   -    & 0.0067 \\
$(\uparrow \uparrow)$     & 0.3181 & 0.3767 & 0.1377 & 0.3252 &   -     \\
\end{tabular}

\end{table}

To quantify the influence of spin and precession effects we compute the mismatch 
between all configurations, see Tab.~\ref{tab:precession}. 
The mismatch is computed from
\begin{equation}
\bar{F} = 1 - \max_{\phi_c,t_c} \frac{(h_1(\phi_c,t_c)|h_2)}{\sqrt{(h_1|h_1),(h_2,h_2)}}
\end{equation}
with $\phi_c,t_c$ being arbitrary phase and time shifts. 
The noise noise-weighted overlap is defined as  
\begin{equation}
 (h_1,h_2) = 4 \Re \int_{f_{\rm min}}^{f_{\rm
 max}} \frac{\tilde{h}_1(f) \tilde{h}_2(f)}{S_n(f)} \text{d} f \ .
\end{equation}
For the one-sided power spectral density of the detector noise $S_n(f)$
we use the \verb#ZERO_DET_high_P# noise curve~\cite{Sn:advLIGO}.
We restrict our analysis to a frequency window of $f \in [450,2048]{\rm Hz}$. 
The chosen lower frequency boundary is slightly above the 
initial GW frequency to avoid imprints of the junk radiation. 
The upper boundary is chosen to be the same as in the analysis of 
the recent BNS detection~\cite{TheLIGOScientific:2017qsa}.

Considering $\iota=0$, we find that the mismatch between spin-aligned configurations and 
the non-spinning configuration is about $3\times 10^{-2}$. 
The mismatch for anti-algined systems is about a factor of $2$ larger which is caused 
by the lower merger frequency of anti-aligned systems. 
The computed mismatch between precessing systems and spinning systems with the same 
effective spin is of the order of $10^{-3}$, which shows that for a face-on 
detection of a BNS hardly any precession effect might be visible from the late inspiral 
phase, see e.g.~\cite{Lindblom:2008cm}. 

For edge-on configurations ($\iota=\pi/2$) we find for $h_+$ similar results as for 
$h_+$ and $h_\times$ for face-on configurations. 
Considering $h_\times$, one finds that systems with the same effective 
spin parameter do result in large mismatches of the order of $\approx 0.3$. 
Interestingly, the mismatch between \Sss and \Snn is about a factor of $6$ 
smaller. However, due to the significantly smaller strain of $h_\times$ for $\iota=\pi/2$
a detection of such a configuration is unlikely.

Overall, we find that in the late inspiral 
precession effects for the employed configurations are small and will most likely 
not be seen in future GW detections with advanced detectors. 
However, studies of the entire inspiral visible in the LIGO-Virgo band needs to be performed for further clarification. 
This will require the creation of waveforms hybridized between our NR waveforms and 
a precessing inspiral model including tidal effects. 
 
\subsubsection{Phase Evolution} 
 
\begin{figure}[t]
\includegraphics[width=.5\textwidth]{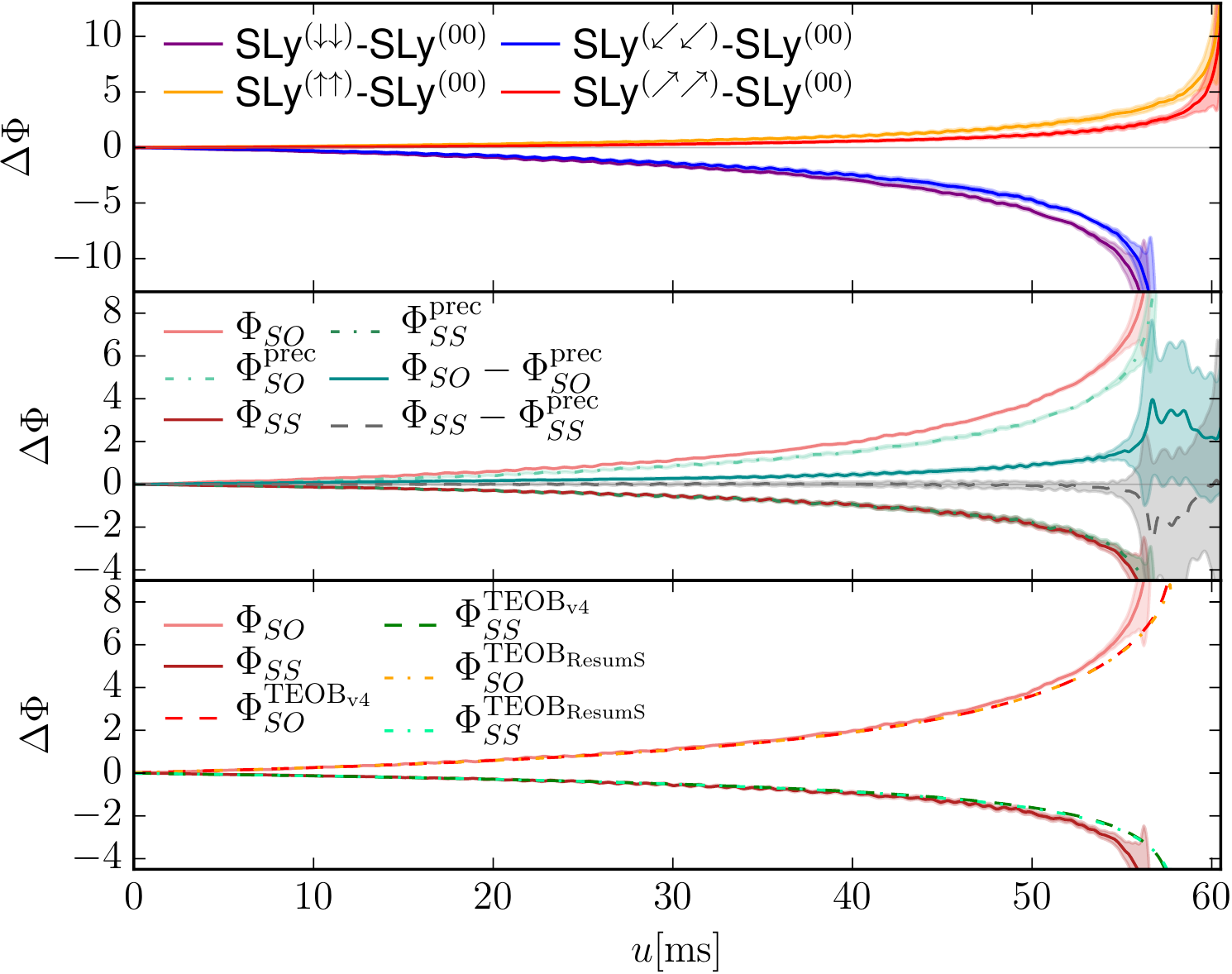}
\caption{Phase differences. 
Top panel: Phase differences for all spinning configurations with 
respect to the non-spinning case. The phase difference is computed 
with respect to $t=0$. The shaded region is estimated by computing the 
phase difference for different resolutions. 
Middle panel: Phase differences between the precessing simulations 
and the spin aligned/anti-aligned configurations.
For all cases we assume an inclination of $\iota=\pi/4$. 
Bottom panel: Estimate of spin-orbit and spin-spin 
contribution to the phase from the NR simulations and 
the EOB approximant TEOBNRv4~\cite{Hinderer:2016eia} and 
TEOBResumS~\cite{Bernuzzi:2014owa,Damour:2014sva,Nagar:2015xqa} restricted to 
the (2,2)-mode. 
Shaded regions correspond to uncertainties assuming error compatible 
to the difference between different resolutions for the individual configurations.}
\label{fig:dephasing}
\end{figure}  
 
In the following we want to discuss shortly the phase differences between 
different configurations. For this purpose we choose an inclination of $\iota=\pi/4$ 
and discuss phase differences with respect to the beginning of the NR simulations
without additional alignment. 

Figure~\ref{fig:dephasing} shows phase differences for the spinning 
configurations with respect to the non-spinning configuration (top panel). 
Visible is an accelerated inspiral due to anti-aligned spin (blue and 
purple curves) and a decelerated inspiral due to aligned spin 
(red and orange curves). These phase differences are again caused mainly by 
the leading order spin-orbit coupling. 
Notice in addition that all curves show small oscillations 
which are caused by the fact that an inclination of 
$\iota = \pi/4$ is chosen. For larger inclinations these oscillations 
further increase.
The opposite holds for smaller inclinations. 

To access the influence of different contribution to the 
phasing~\footnote{For a more quantitative discussion also time correction due to 
spin and tidal effects need to be included. 
We postpone this kind of analysis and an analysis of the frequency domain 
phase to future work with simulations at higher resolution and the improved numerical methods
of~\cite{Bernuzzi:2016pie}.}, 
we consider, as for the binding energy, different linear combinations of 
different numerical simulations, in particular we consider
the spin-orbit contribution: 
\begin{eqnarray}
\Phi_{SO}            & = & (\Phi[{\rm SLy}^{(\uparrow \uparrow)}] - 
\Phi[ {\rm SLy}^{(\downarrow \downarrow)}])/2, \\
\Phi_{SO}^{\rm prec}  & = & (\Phi[{\rm SLy}^{(\nearrow \nearrow)}] - 
\Phi[ {\rm SLy}^{(\swarrow \swarrow)}])/2,
\end{eqnarray}
and spin-spin contribution: 
\begin{small}
\begin{eqnarray}
\Phi_{SS}            & = & (\Phi[{\rm SLy}^{(\uparrow \uparrow)}] + 
\Phi[{\rm SLy}^{(\downarrow \downarrow)}])/2 - \Phi[{\rm SLy}^{(00)}], \\
\Phi_{SS}^{\rm prec}  & = & (\Phi[{\rm SLy}^{(\nearrow \nearrow)}] + 
\Phi[{\rm SLy}^{(\swarrow \swarrow)}])/2 - \Phi[{\rm SLy}^{(00)}]. 
\end{eqnarray}
\end{small}
The middle panel of Fig.~\ref{fig:dephasing} shows our main results. 
The spin-orbit contribution dominates. 
Although the precessing configurations and configurations with aligned/anti-aligned 
are chosen such that the leading order spin-orbit contribution is equal, 
one sees small differences during the inspiral. 
The spin-spin effect is significantly smaller than the spin-orbit effect
and interestingly is almost identical for the precessing simulations 
and simulations with aligned/anti-aligned spin. 
We surmise that the reason for this effect is that, although 
the spin magnitudes differ, terms proportional to $\propto (\vv{S}_i \cdot \vv{L})$
present in the spin-spin contributions are similar. 

Finally, as a consistency check, we compare $\Phi_{SO}$ and $\Phi_{SS}$ 
with results obtained with a spin-aligned tidal effective-one-body (EOB) approximant
(bottom panel of Fig.~\ref{fig:dephasing}).  
As tidal EOB approximants we use the TEOBNRv4 model of the LIGO Algorithm 
Library~\cite{Hinderer:2016eia} and the TEOBResumS model~\cite{Nagarinprep}.
The phases of the tidal EOB approximants are extracted solely from the dominant (2,2)-mode.
Both EOB predictions are almost indistinguishable and 
overall one finds very good agreement between the tidal EOB models and 
NR results. This shows the accuracy of our numerical simulations and the 
validity of current state-of-the-art waveform models. 

\subsection{Postmerger evolution} 
 
\begin{figure}[t]
\includegraphics[width=.5\textwidth]{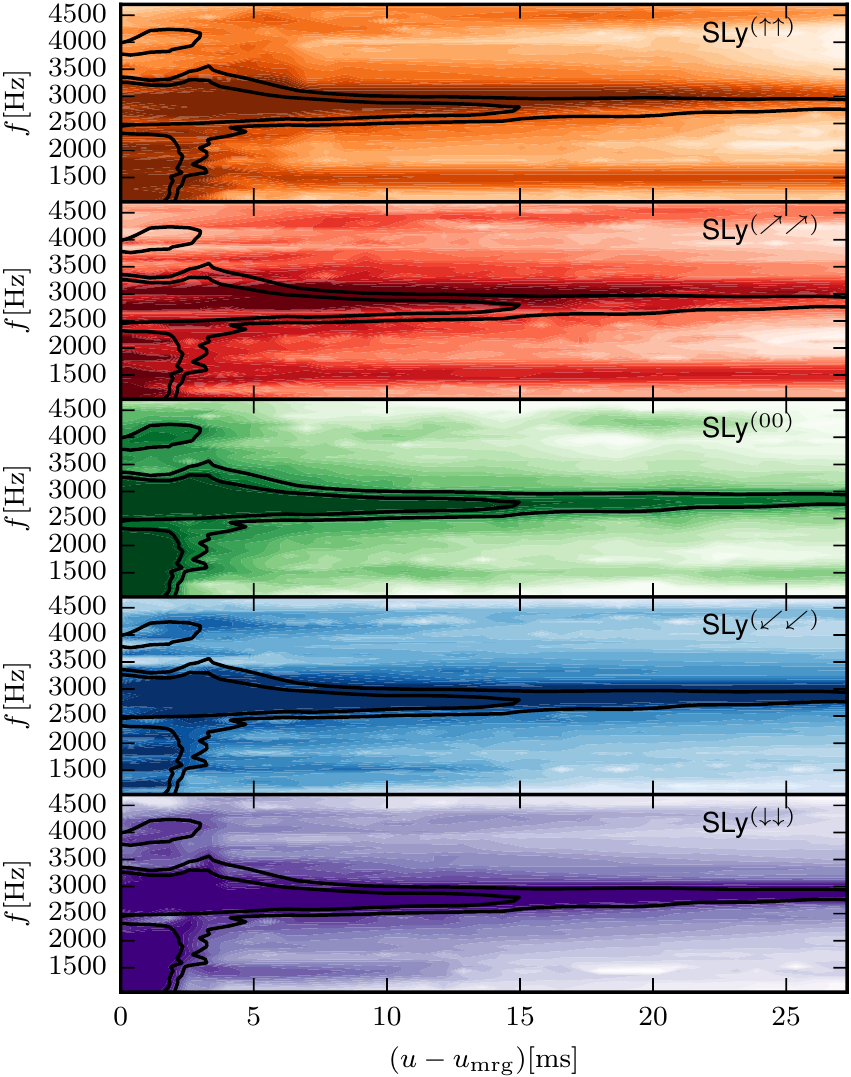}
\caption{Spectrogram for the all configurations. 
         We shift the beginning of the spectrogram to the 
         merger time (peak in the gravitational wave strain). 
         We include as a black contour the spectrogram of \Soo 
         to all plots. All plots assume an inclination angle 
         of $\iota=\pi/4$.}
\label{fig:spec}
\end{figure}  
 
To access the postmerger GW signal, we compute spectrograms following~\cite{Dietrich:2016hky}, 
see also~\cite{Bernuzzi:2013rza,Clark:2015zxa,Rezzolla:2016nxn}. 
Figure~\ref{fig:spec} shows the postmerger spectrograms for all configurations under 
the assumption of $\iota = \pi/4$.
We start by discussing the non-spinning configuration \Soo in more detail and later 
point out differences caused by the intrinsic rotation of the stars. 

The middle panel (green) of Fig.~\ref{fig:spec} shows \Soo. 
Most prominent is the $f_2$-peak frequency $f_2\approx 2800 {\rm Hz}$. 
One sees that the amplitude of the peak is decreasing while its frequency increases. 
The frequency increase is caused by the increasing compactness of the star due to the emission 
of angular momentum in terms of GWs. 
In addition to the $f_2$ peak, additional side peaks and frequencies are visible. 
Those correspond to the $m=1$ and $m=3$
emission at about $f_1 \approx 1500 {\rm Hz}$ and $f_3 \approx 4200 {\rm Hz}$, respectively. 
These peaks are harmonic to the $f_2$ frequency.
As pointed out in e.g.~\cite{Takami:2014zpa,Bauswein:2014qla,Rezzolla:2016nxn} around the moment of merger 
additional frequency peaks are present, see e.g.~the peak at $f\approx 4000 {\rm Hz}$. 

Considering the difference between \Soo and \Suu, \Snn, we find that the additional peak around merger 
at $f\approx 4000 {\rm Hz}$ is not present. However, the $f_1$ peak is significantly more pronounced 
during the postmerger evolution for 
systems with aligned spin. Most importantly, a frequency shift to higher frequencies 
of about $200 {\rm Hz}$ is present. 
Those shifts are of particular importance in case quasi-universal 
relations~\cite{Bauswein:2014qla,Bauswein:2015yca,Takami:2014tva,Takami:2014zpa,Bernuzzi:2015rla} are employed to 
constrain the NS radius from the postmerger 
signal~\cite{Bose:2017jvk,Chatziioannou:2017ixj}. 
Existing relations have been derived for irrotational binaries and might contain systematic 
biases for spinning systems. 

For configurations with anti-aligned spin we find that the additional peaks around merger are visible, 
see in particular the \Sdd configuration. Contrary, the $f_1$ peak is less pronounced after the merger. 
Considering the $f_2$-peak frequency, we find that \Sss and \Sdd have a similar evolution
as \Soo. 

We do not find noticeable differences between the precessing systems and systems with 
the same effective spin, but purely aligned or antialigned spin. Thus, we 
conclude that at least from our simulations we can not impose constraints on precession effects 
during the inspiral from the postmerger GW signal. 

\section{Ejecta} 
\label{sec:ejecta}

\begin{table}[t]
  \centering
  \caption{Ejecta mass and kinetic energy extracted form our simulations given in geometric units.}
  \label{tab:ejecta}
  \begin{tabular}{l|ccccc} 
  \hline
             & \Suu            &    \Snn          &    \Soo           &    \Sss         &   \Sdd \\ 
    \hline
 $M_{\rm ej}$& $5\times10^{-3}$& $4\times10^{-3}$ & $7\times10^{-3}$  & $7\times10^{-3}$& $8\times10^{-3}$ \\
 $T_{\rm ej}$& $1\times10^{-4}$& $6\times10^{-5}$ & $2\times10^{-4}$  & $2\times10^{-4}$& $2\times10^{-4}$ \\
    \hline
  \end{tabular}
\end{table}

In addition to the GW signal emitted from BNS coalescence, 
possible electromagnetic signals give important information about 
the binary parameters. 
To predict the properties of the 
kilonovae~\cite{Metzger:2011bv,Tanvir:2013pia,Yang:2015pha,Metzger:2016pju,Tanaka:2016sbx} 
and radio flares~\cite{Nakar:2011cw}
one has to know the amount of ejected mass, its geometry, velocity, and composition. 
In this article we restrict our investigations to dynamical ejected material, which becomes 
unbound around the time of merger. 

In Tab.~\ref{tab:ejecta}, we give estimates of the masses as well as about the kinetic energy 
of the ejecta. The uncertainty caused by resolution effects are of the order of $\sim 20\%$, 
however, larger systematic uncertainties are present due to the artificial atmosphere 
employed for the evolution~\cite{Dietrich:2016hky}. 
Consequently we want to restrict ourselves to a qualitative 
discussion. 

The mechanisms for dynamical ejecta are either torque in the tidal 
tail of the stars or shocks created during the merger. 
In Ref.~\cite{Dietrich:2016lyp} was shown that for 
aligned spin systems torque inside the tidal tail increases. 
This leads to more massive ejecta in cases in which torque driven ejecta dominate. 
Here we find that the ejecta mass increases when the NS spin is anti-aligned to the 
orbital angular momentum. 
This indicates that the dominant ejecta mechanism are shocks produced during the merger 
of the two stars, see also~\cite{Kastaun:2016elu}. 
A similar trend is observable for the kinetic energy, 
where overall non-spinning or anti-aligned configurations
have ejecta with largest kinetic energy, which will result in 
larger fluencies of possible radio flares. 

We find only a small imprint of precession on the amount of ejected material
and differences between precessing systems and systems with the same effective spin 
are within the uncertainty of our data. 
Due to the small precession angle of $\approx 6^\circ$ of the orbital angular momentum, 
the ejecta morphology is similar for precessing and non-precessing systems. 
However, we suggest that for larger spin values or stiffer EOSs precession 
effects might also be observable. Studying such configurations is left for the future.  

\section{Conclusion}
\label{sec:conclusion}
In this article we studied the effect of spin precession in binary
neutron star merger simulations. 
We considered two precessing configurations, two configurations 
with aligned/anti-aligned spin, and one non-spinning case. 
All configurations have been simulated with three different resolutions to 
assess the uncertainties of our simulations. 
We find that state-of-the-art numerical relativity simulations
are capable to simulate precessing neutron star systems 
and that spin and precession effects are well resolved. 

To interpret the inspiral dynamics, we presented the precession cones 
of the individual spins and orbital angular momentum. 
Although the individual spin of NSs in binaries is not 
defined in general relativity a simple quasi-local spin estimate gives 
results close to PN predictions. 

To quantify the inspiral dynamics we constructed binding energy vs.~specific 
angular momentum curves, finding that precession effects are hardly visible
in the conservative dynamics and that the main effect on the 
binary evolution is, as expected, the spin-orbit interaction. 

Considering the effect of precession on the GW signal, we found that 
for edge-on systems precession effect in the late inspiral are clearly 
visible. However, those systems are harder to detect 
due to the smaller observable gravitational wave strain for such
inclinations.
In contrast for face-on systems precession effects seem to be hardly 
detectable. 
Similarly, while the postmerger waveform shows a clear imprint of 
spin effects in terms of a shift of the main emission frequency of 
about $200 {\rm Hz}$, no noticeable imprint of precession effects is visible. 
Similarly, in addition to the gravitational wave emission,
possible electromagnetic counterparts, mostly triggered by the 
ejected material, are also unlikely to show noticeable precession effects for the considered configurations.  

Further work is needed to quantify the imprint of precession not only during the last 
milliseconds of the binary neutron star coalescence, but 
for the full inspiral visible by current gravitational wave detectors 
already seconds before the actual merger.  
Additionally, simulations of other configurations with different 
EOSs, total masses, and mass ratios employing higher resolution 
are ongoing to further investigate precession effects 
in a larger region of the binary neutron star parameter space.

\begin{acknowledgments}

  It is a great pleasure to thank S.~Ossokine for 
  very helpful and fruitful discussions and 
  support with \texttt{GWframes}.
  We are also thankful to K.~Kawaguchi for computing the 
  TEOBNRv4 and A.~Nagar for computing the TEOBResumS waveforms 
  used for Fig.~\ref{fig:dephasing}.
  
  S.B.~acknowledges support by the European Union's H2020 under ERC Starting Grant, grant
  agreement no. BinGraSp-714626. 
  W.T.~was supported by the National Science Foundation under grants
  PHY-1305387 and PHY-1707227.
  M.U.~is supported by Funda\c{c}\~ao de Amparo \`a Pesquisa do Estado de
  S\~ao Paulo (FAPESP) under the process 2017/02139-7.
  Computations were performed on SuperMUC at the LRZ (Munich) under 
  the project number pr48pu, Jureca (J\"ulich) 
  under the project number HPO21, Stampede 
  (Texas, XSEDE allocation - TG-PHY140019), 
  Marconi (CINECA) under ISCRA-B the project number HP10BMAB71, 
  under PRACE allocation from THIER0 call 14th, 
  and on the Hydra and Draco clusters of 
  the Max Planck Computing and Data Facility. 
\end{acknowledgments}


\bibliography{paper20171208.bbl}

\end{document}